# Single power-law rheology of crowded cytoplasm in living cells

H. Ebata,[1] K. Nishizawa,[2] F.A.S. van Esterik,[1] Y. Tao,[1] S. Inokuchi,[1] H. Ise,[3] and D. Mizuno,[1*]

[1]Department of Physics, Kyushu University, 819-0395 Fukuoka, Japan.
[2]Department of Applied Physics, Graduate School of Engineering, Tohoku University, 980-8579 Sendai, Japan.
[3]Institute for Materials Chemistry and Engineering, Kyushu University, 819-0395 Fukuoka, Japan.

\* Correspondence: mizuno@phys.kyushu-u.ac.jp; Tel.: +81-92-802-4092

**Abstract:**

Cytoplasmic viscoelasticity is crucial for various intracellular processes. However, the dynamic shear modulus, $G(\omega)$, has been reported to vary considerably, often without consistent patterns or rules, even within the same cell. Thus, uncovering the physical basis of cytoplasmic rheology, and whether any universal feature exists, remains a major challenge. Here, we employed microrheology with a 3D feedback technique to minimize artifacts such as laser phototoxicity and examined cytoplasmic viscoelasticity across varied mechanical environments, cell types, and cytoskeletal disruptions. Unlike previous studies, a single power-law rheology $G(\omega) \propto (-i\omega)^{0.5}$ was observed over a broad frequency range for all conditions except ATP depletion. While the vimentin cytoskeleton significantly contributed to steady shear viscosity measured by pulling a particle over large distances, cytoskeletal disruptions had only a minor effect on locally measured viscoelasticity. These findings demonstrate that molecular crowding governs the observed universality, providing a framework to systematically investigate cytoplasmic mechanics across diverse cellular contexts.

## INTRODUCTION

Dynamic mechanical properties of the cell interior are fundamental for various physiological processes, such as biochemical reactions[1], transport of organelles[2,3], deformation of a cell body, and cell migration[4,5]. Elastic and viscous forces from the surrounding cytoplasm govern the movement of biomacromolecules and organelles within a cell. Therefore, the material properties of the cell cytoplasm, i.e., diffusion coefficient, viscosity, and elasticity, play a key role in maintaining cell functions.

The cell interior is composed of active soft materials with properties intermediate between those of an elastic solid and a viscous fluid[6,7]. The dynamic mechanical properties of a cell, represented by the complex shear modulus $G = G' - iG''$, depends on the angular frequency $\omega = 2\pi f$ of the periodically applied force[8]. Here, $G'(\omega)$ and $G''(\omega)$ represent the storage and loss moduli, respectively. The complex shear modulus of cells was reported to exhibit a power-law dependence on frequency[9,10], $G \propto (-i\omega)^\alpha$. This differs from traditional viscoelastic materials such as Kelvin-Voigt solids and Maxwell fluids which show a single Debye relaxation. The power-law exponent $\alpha$ of the cell interior typically increases as the frequency shifts from low ($f \sim 0.1 - 10$ Hz) to high ($f \sim 100 - 10000$ Hz) frequencies[7,11-13]. Notably, the coefficients and exponents of the power-law vary across different cells[7,10,14], at different points in the cell cycle[11], and in response to surrounding micromechanical environments[15,16], making it challenging to systematically evaluate cell rheology.

Many previous studies measuring cellular viscoelasticity have used experimental methods that primarily reflect the property of cytoskeleton, such as actin, vimentin and microtubule networks[7,11,17]. Cytoskeletal organization can vary significantly over time[18,19], depending on the intracellular location[20,21], and extracellular condition[22-24], which accounts for the broad variability in measured viscoelasticity (see Fig. 1). In contrast, when macromolecular crowding dominates over the cytoskeleton, the cytoplasm may exhibit a single power-law rheology as $G \propto (-i\omega)^{0.5}$ [6,25]. However, artifacts such as laser phototoxicity have affected prior intracellular measurements, leaving it uncertain whether cytoplasmic rheology universally reflects features arising from crowding.

Investigating the local mechanics within the deeper regions of a cell has long been a challenge. Some prior studies attempted to address this by employing passive microrheology (PMR)[7,26,27], which observes the spontaneous fluctuations of probe particles within a specimen[28-30]. Under thermal equilibrium, it is possible to obtain the complex shear modulus of the surrounding specimen from the probe particle displacements, using the fluctuation-dissipation theorem (FDT) and the generalized Stokes relation[31]. However, because the cell interior is far from equilibrium, the FDT is violated[11,32].

In living cytoplasm, mechano-enzymes (e.g., motor proteins) generate forces utilizing the energy released through hydrolyzing nucleoside triphosphates (ATP). This drives dynamic processes, such as organelle transport, protoplasmic flow, and actomyosin fiber contraction. Nonequilibrium fluctuations induced by these processes[33,34] contribute to the movements of probe particles measured by PMR whereas the FDT accounts only for thermal contributions. As a result, when nonequilibrium fluctuations are not negligible, PMR underestimates the complex shear modulus.

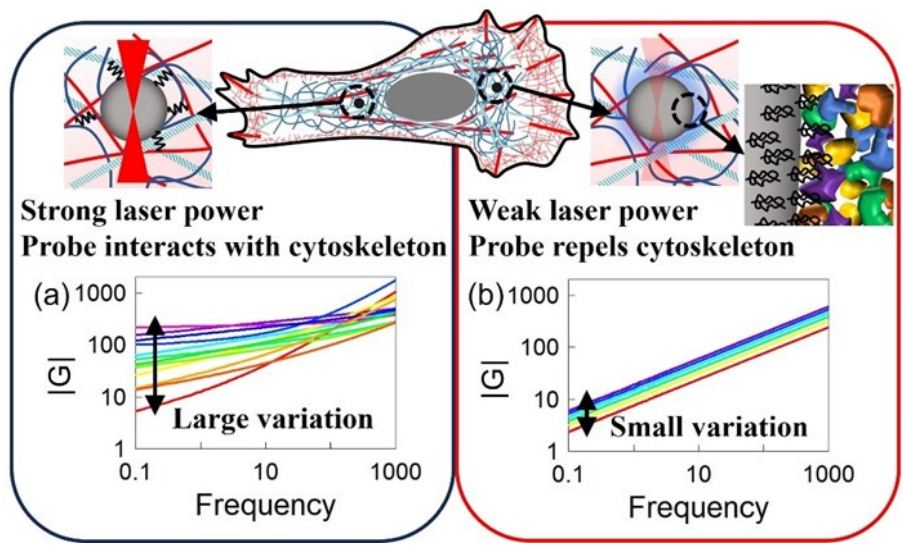

**Figure 1. Schematic illustration of intracellular viscoelasticity**

(a) Conventional measurement using laser interferometry. Measured cell viscoelasticity strongly depended on intracellular locations, micromechanical environments, and cell cycles, possibly relating to cytoskeletal development. (b) The single power-law rheology $G = G_1(-i\omega)^{0.5}$ observed in this study. The magnitude of cytoplasmic viscoelasticity is almost independent of the state of cytoskeleton. (a, b) Colored lines schematically represent the variation of measured complex modulus under different conditions.

Instead of PMR, intracellular rheology can be directly assessed by measuring the response of a probe particle inside a cell to an applied force (active microrheology: AMR[35]). In AMR, an optical trap is used to exert an external force on a probe particle (~ μm) dispersed inside a cell. Its displacement response is measured using the back-focal-plane laser interferometry (BFPI) technique[11,14,31,36]. However, this approach presents challenges when attempting to apply a force with high spatiotemporal resolution over extended periods. Intense intracellular fluctuations, such as protoplasmic flow, can destabilize an optical trap, and sometimes even causing optical trapping to fail altogether. Even when a probe particle inside a cell is successfully trapped, precise control of the force applied to the probe particle remains challenging. The optical-trapping force depends on the relative position between the laser and the probe particle; thus, if the probe particle fluctuates, the force also fluctuates stochastically. Conventionally, strong laser power has been used to stabilize the trap. However, the strong potential generated by a strong laser suppresses both the thermal motion of the particles and their responses to applied forces, leading to significant errors in estimating inherent fluctuations and responses[14]. Furthermore, excessive laser power alters cytoplasmic mechanics[37] due to phototoxic effects on the cytoplasm[38-40], introducing additional variability in the measured mechanical properties (see Fig. 1).

Here, we ask whether a consistent viscoelastic behavior can be observed in the cytoplasm of living cells when experimental artifacts are eliminated, and what factors - such as cytoskeletal structure, metabolic activity, mechanical environment, or cell type - predominantly determine intracellular mechanics. In our previous study, we developed a microrheology technique that combines optical trapping with a 3D position feedback system (feedback AMR)[14,37]. This system enables prolonged measurements with vigorously fluctuating probe particles at laser intensities much lower than those used in conventional optical-trapping measurements (> 5 mW)[36], although the technology's full potential has not yet been explored. In this study, we refined the method to perform AMR measurements in cells using a minimal laser power (~ 0.5 mW)[14,37]. This unique capability of feedback AMR reduced potential artifacts, such as estimation errors of the response function and phototoxic effects, to negligible levels.

By performing feedback AMR under various mechanical environments, with cytoskeletal inhibitors, and in different cell types, we identified universal dynamic mechanical properties of living cytoplasm. Cell-specific or context-specific intracellular rheology reported in prior studies largely vanished by minimizing photodamage and probe-cytoskeleton coupling. Under all conditions tested, except for ATP depletion, the complex shear modulus consistently followed a single power-law, $G = G_1(-i\omega)^{0.5}$, over a broad frequency range (0.1 – $10^5$ Hz). Neither cytoskeletal inhibition nor changes in micromechanical environments did not significantly alter the power-law exponent and only slightly affected $G_1$, indicating that our measurements primarily reflect the rheology of the cytoplasm. In this context, 'cytoplasm' describes the intracellular medium excluding contributions from the cytoskeleton, commonly described as crowded cytoplasm[41]. The observed exponent of 0.5 closely resembles that of critical jamming rheology[42], suggesting that the cytoplasm behaves as a densely packed glassy material.

While the observed power-law rheology was minimally affected by the cytoskeleton in the range of frequencies measured, the shear viscosity of cells at nearly zero frequency was significantly influenced by the vimentin network[43]. This finding was made by long-term (~1h) constant-force pulling of probe particles, probing large-scale (~ μm) cytoplasmic properties. In contrast, AMR uses probe particles oscillated with small amplitudes (< nm), and cytoskeletal filaments are sterically excluded from the probe surface. The classical view emphasizing the cytoskeletal role on cell mechanics might seem to contradict our finding that cytoplasmic crowding governs the universal intracellular rheology. However, these views are reconciled by recognizing their distinct spatiotemporal regimes: cytoskeletal filaments, especially vimentin, regulate slow, large-scale mechanics. In contrast, the universal power-law rheology observed across diverse conditions arises from molecular crowding at smaller length scales—a regime in which the cytoskeleton has only a marginal influence.

## RESULTS

### Single power-law rheology of HeLa cells

AMR was conducted to measure intracellular viscoelasticity without requiring strong optical trapping (see Fig. 2 (a) and Methods for details). To stabilize the trap with a weak laser, the system dynamically adjusted the laser focus to track the center of the rapidly moving probe particle using feedback control of the piezo-mechanical stage (Fig. 2 (b)). To determine the complex modulus, a weak oscillatory force was applied, inducing small amplitude oscillations in the probe particle's position (Fig. 2 (c)). Figure 3 (a) shows the median value of the complex shear modulus of HeLa cells (n = 12) measured with a low laser power (~ 0.5 mW). Circles and squares represent the real ($G'$) and imaginary ($G''$)

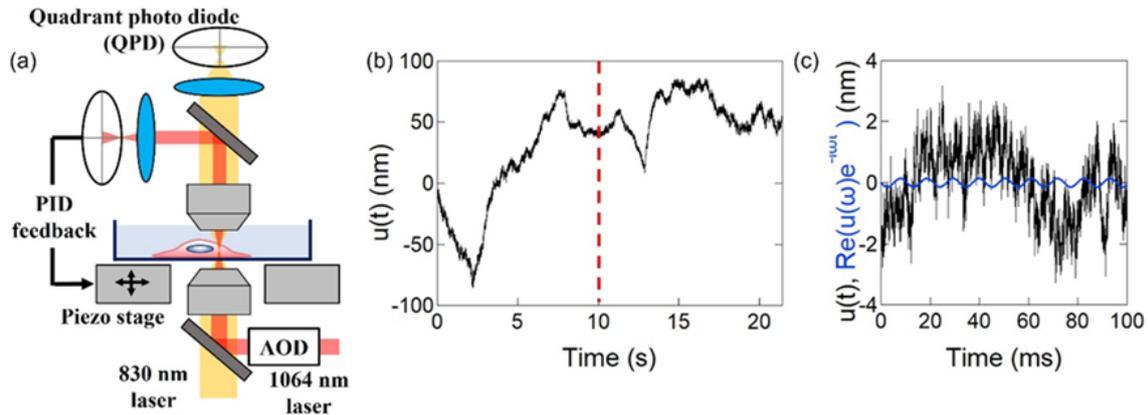

**Figure 2. The setup of stage feedback active microrheology (AMR)**

(a) A schematic illustration of the stage feedback microrheology setup. A probe laser (830 nm) was used to measure the particle position $u_{QPD}$ by a quadrant photodiode (QPD). An oscillatory force was applied by a drive laser (1064 nm) with an acousto-optic deflector (AOD). (b) Representative probe trajectory $u(t)$ recorded during feedback AMR inside a HeLa cell. The lasers didn't disturb spontaneous fluctuation of the probe particle because the laser focus continuously tracked the center of the probe particle. (c) Enlarged view around the red dashed line in Fig. 2 (b). The time averaged value of $u(t)$ was subtracted. Blue curve $\mathrm{Re}\left(u(\omega)e^{-i\omega t}\right)$ denotes oscillation induced by the applied force, which is much smaller than the spontaneous fluctuation of the probe. Oscillation (blue curve) was detected by using a lock-in amplifier.

parts of the complex shear modulus, $G = G' - iG''$, respectively. Since the distributions of the complex modulus deviate from a Gaussian distribution[7], the median values of the experimental results are plotted with error bars indicating the interquartile range (25% to 75% quantiles). The complex modulus followed a single power law $G = G_1(-i\omega)^{0.5}$ across a broad frequency range ($10^{-1} - 10^5$ Hz). In our previous work[37], $G'(\omega)$ deviated from the single power-law response, $(-i\omega)^{0.5}$, displaying an elasticity plateau ($\sim 10$ Pa) at low frequencies ($\leq 1$ Hz). As seen in Fig. 3 (b), this elasticity plateau was attributed to laser-

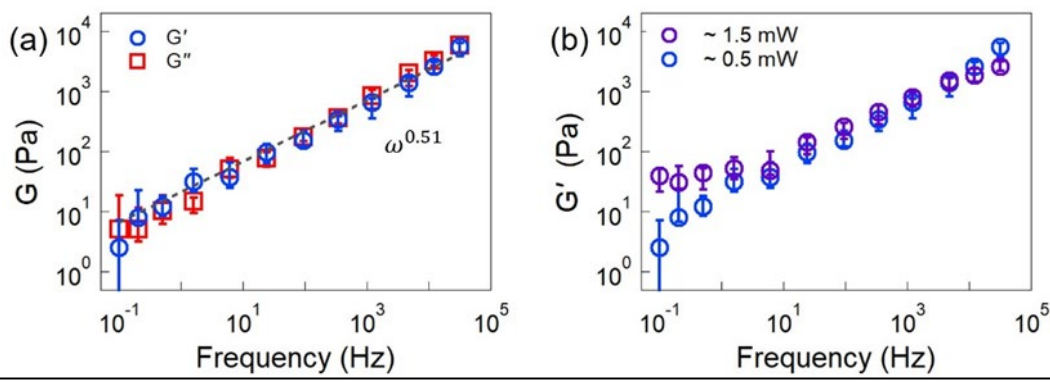

**Figure 3. Single power-law rheology of HeLa cells**

(a) Viscoelastic moduli $G'(\omega)$ (bule circle) and $G''(\omega)$ (red square) of HeLa cells in a confluent monolayer formed on a glass substrate (n = 12). Symbols represent the median values, and bars indicate the interquartile range (25% to 75% quantiles). The broken solid line shows the fitting of $G = G_1(-i\omega)^n$, where $n = 0.51 \pm 0.04$. (b) Elasticity plateau was induced by Laser-induced photodamage. The power of the drive laser was approximately 0.5 mW (blue circle) and 1.5 mW (purple circle), respectively. Melamine particles ($2a = 1$ μm) were used as a probe.

induced photo-damage. Specifically, approximately 1.5 mW of laser power caused a plateau in $G'(\omega)$ at frequencies below 5 Hz, while reducing the laser power to $\sim 0.5$ mW eliminated this hardening effect. Therefore, in this study, AMR experiments were performed using a weak laser power ($\sim 0.5$ mW) unless otherwise stated.

Remarkably, our measurements exhibited limited variability in viscoelasticity among cells. Assuming the absence of an elastic plateau, fitting of the single power-law model $G = G_1(-i\omega)^{0.5}$ indicated only marginal differences in the viscoelastic coefficient ($G_1 = 12.0 \pm 6.6$) and the exponent ($n = 0.51 \pm 0.04$). These results suggest that the mechanical properties of the crowded cytoplasm are relatively homogeneous compared to the heterogeneity stemming from cytoskeletal structures[44].

**Limited contribution of cytoskeleton to intracellular microrheology**

To investigate how cytoskeletal components contribute to cytoplasmic rheology, we measured the viscoelasticity of HeLa cells after disrupting actin, microtubule, and vimentin. Actin polymerization was inhibited by treating cells with cytochalasin D at 2.5 μg/ml for $\sim$ 1 h, which induced noticeable morphological changes such as the loss of cell-cell contacts and a weaker attachment to the bottom substrate. The cells became rounded with branched protrusions (Fig. 4 (a)). Despite these morphological alterations, the viscoelastic relationship $G = G_1(-i\omega)^{0.5}$ remained unchanged, and the coefficient $G_1$ showed no significant difference from untreated cells (Fig. 4 (b), (c))[37].

For microtubule disruption, cells were treated with 3 μg/ml nocodazole for $\sim$ 2 h. This treatment caused moderate but noticeable morphological changes: the edge of lamellipodia retracted, and the cytoplasmic contents concentrated around the nucleus (Fig. 4 (a)). Small gaps appeared between cells although they remained attached to the bottom substrate. As with actin disruption, microtubule disruption had only a minor effect on the cytoplasmic viscoelasticity (Fig. 4 (b), (c)).

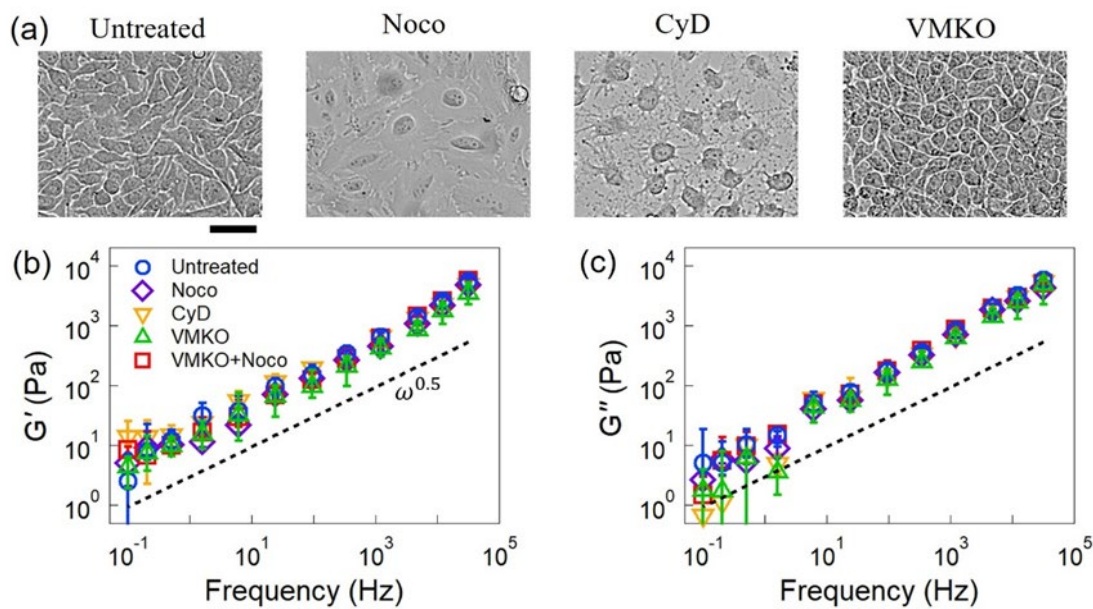

**Figure 4. Limited contribution of cytoskeleton to intracellular microrheology**
(a) A microscopic image of untreated HeLa cells, HeLa cells treated by nocodazole (Noco), cytochalasin D (CyD) and vimentin-knockout (VMKO) HeLa cells. (b) $G'(\omega)$ and (c) $G''(\omega)$ of HeLa cells with various cytoskeletal disruptions. Blue circle: untreated HeLa cells (n = 12). Purple diamond: HeLa cells with microtubule disruption (n = 7). Yellow down triangle: HeLa cells with actin disruption (n = 8). Green up triangle: vimentin-knockout HeLa cells (n = 14). Red square: vimentin-knockout HeLa cells with microtubule disruption (n = 4). The broken line shows the power law $\propto \omega^{0.5}$. The power-law rheology remained unaffected by cytoskeletal disruptions.

Vimentin-knockout HeLa cells exhibited a smaller and more elongated phenotype compared to typical HeLa cells (see supplementary material, Fig. S2). Although vimentin is highly expressed in normal HeLa cells[45], its disruption only slightly decreased cytoplasmic viscoelasticity (Fig. 4 (b), (c)), contrasting with a previous study[17]. Further disruption of the microtubule network in vimentin-knockout HeLa cells led to a moderate increase in the complex modulus. Overall, cytoskeletal disruptions have only minor effects on the viscoelasticity of crowded cytoplasm.

**Effect of the mechanical environment on cytoplasmic viscoelasticity**

To investigate the impact of the mechanical environment, AMR was performed in HeLa cells under different conditions of cell density and substrate stiffness. Specifically, three experimental conditions were examined: isolated and confluent HeLa cells cultured on a glass substrate, and isolated cells cultured on soft polyacrylamide (PAA) gel substrates (Fig. 5 (a)).

As shown in Fig. 5 (b), (c), the single power-law viscoelasticity $G = G_1(-i\omega)^n$ was hardly affected by both cell density and substrate stiffness. For cells cultured on glass substrates, $G_1 = 10.0 \pm 6.1$ and $n = 0.51 \pm 0.04$ obtained in isolated HeLa cells were not significantly different from those in confluent cells. Cells grown to confluency became thicker with heights comparable to their width, while isolated cells exhibited a well-adhered

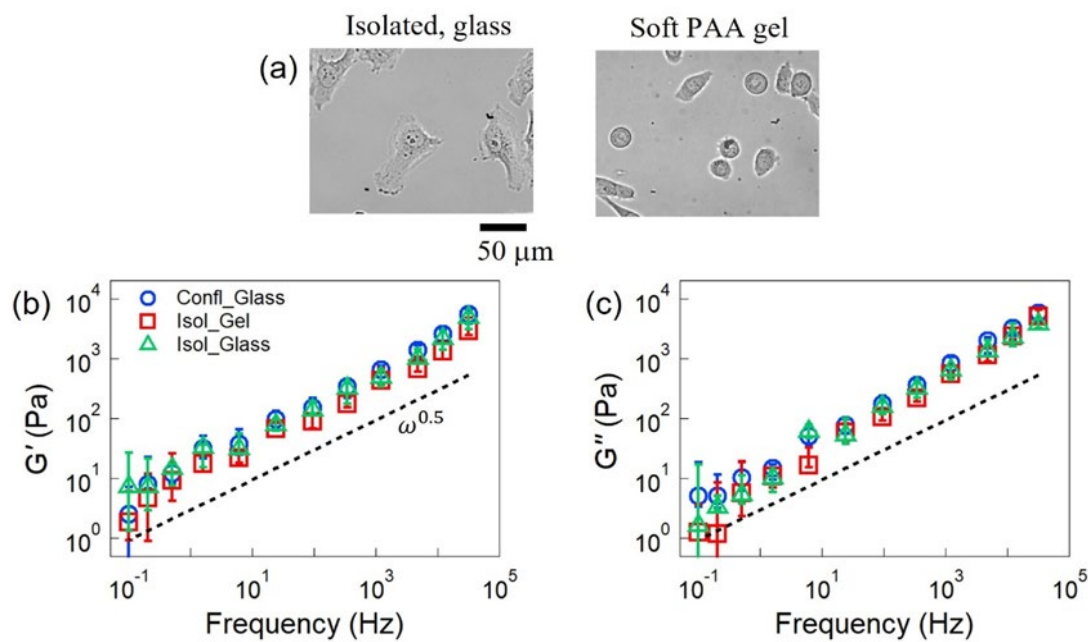

**Figure 5. Effect of the mechanical environment on cytoplasmic viscoelasticity**

(a) A microscopic image of isolated HeLa cells on a glass substrate (left) and HeLa cells on a soft PAA gel (right). (b) $G'(\omega)$ and (c) $G''(\omega)$ of HeLa cells in a confluent monolayer formed on a glass substrate (blue circle, n = 12), a single isolated cell on a glass substrate (green triangle, n = 6), and isolated cells on a soft PAA gel with an elasticity of ~ 200 Pa (red square, n = 9). The broken line shows the power law $\propto \omega^{0.5}$. The power-law rheology remained unaffected by micromechanical environment.

flattened morphology. In isolated cells, AMR measurements were conducted in regions excluding pseudopodia, while in confluent cells, probe particles that were not too close to the membrane or nucleus were used. Under these conditions, cell density had no significant effect on cytoplasmic viscoelasticity.

Adhesion of HeLa cells to a soft substrate is weak, which is known to affect cell morphology. As seen in Fig. 5 (a), HeLa cells on soft gels were more rounded compared to those on glass substrate. While spread cells on a glass displayed migration through extension and contraction of their body, spherical cells on soft substrates remained relatively stationary[46]. Despite these morphological and migratory differences, substrate stiffness had no significant impact on the complex shear modulus $G$ (red square and green triangle in Fig. 5 (b) and (c)). Thus, changes in the micromechanical environment have a limited effect on the rheology of crowded cytoplasm.

In HeLa cells cultured on soft gels ($\sim$ 1 kPa), actin, vimentin and microtubule network become sparse except at the periphery, whereas on glass substrates, these cytoskeletal structures develop more densely[47,48]. Similarly, in confluent monolayers, the actin cytoskeleton primarily localizes around the cell borders and at the basal plane, with sparse expression deeper within the cell[49]. Although these differences in cytoskeletal organization were observed as expected, changes in the micromechanical environment induced no variations in cytoplasmic viscoelasticity, as shown in Fig. 5. These results coincide with the limited effect of cytoskeletal disruption reported in the previous section.

**Cytoplasmic rheology depends on metabolic activity**

Unlike previous works[11,17], our intracellular microrheology experiments primarily reflect the viscoelasticity of the cytoplasm in the regions between the cytoskeletal filaments. Therefore, cytoskeletal development and associated properties, and behaviors of cells have limited impact on the intracellular rheology measured in this study. On the other hand, we found that metabolic activity plays a crucial role in regulating the storage modulus $G'(\omega)$ at low frequencies[37]. As shown in Fig. 6 (a), (b), when intracellular ATP was depleted in HeLa cells, $G'(\omega)$ exhibited an elasticity plateau below 10 Hz, while $G''(\omega)$ did not change appreciably. While similar tendency was reported in our previous study[37], those data were obtained using stronger laser, possibly inducing laser-induced photodamage to the cytoplasm. In this study, we used a weaker laser ($\sim$ 0.5 mW) to completely avoid such phototoxic effects. This technical improvement successfully eliminated artifacts in untreated HeLa cells, resulting in an elasticity plateau in $G'(\omega)$, observed only in ATP-depleted cells.

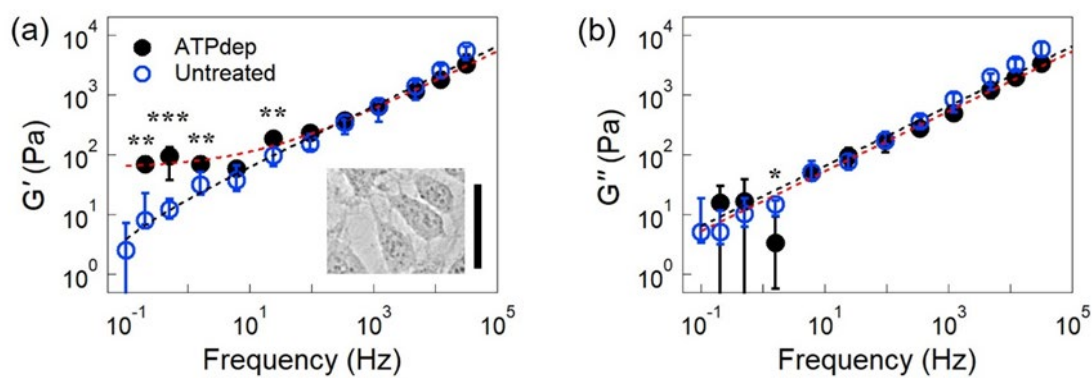

**Figure 6. Cytoplasmic rheology depends on metabolic activity**
(a) $G'(\omega)$ and (b) $G''(\omega)$ of ATP-depleted (ATPdep) HeLa cells (solid black circle, n = 7) and untreated cells (blue circles, n = 12). (Inset) A microscopic image of ATP-depleted HeLa cells. Scale bar: 50 μm. Dashed line: fitted curve by $G = G_0 + G_1(-i\omega)^{0.5}$. The Mann–Whitney U test was used to calculate the P value. *: P < 0.05. **: P < 0.01. ***: P < 0.001.

**Quantitative comparison of cytoplasmic viscoelasticity under various conditions**

To quantitatively evaluate the effects of cytoskeletal disruption, micromechanical microenvironments, and ATP depletion, the complex moduli were fitted using $G = G_0 + G_1(-i\omega)^{0.5}$. As shown in Fig. 7 (a), the elasticity plateau $G_0$ appeared only when ATP was depleted. In contrast, the power-law component of viscoelasticity, represented by $G_1$, did not show significant differences under various conditions compared to confluent HeLa cells on glass substrates (Fig. 7 (b)). However, $G_1$ exhibited a weak but notable deviation from the ordinary value under certain conditions. Specifically, $G_1$ in vimentin-knockout cells tended to be lower than in untreated cells with P = 0.068.

The emergence of a non-zero elasticity plateau $G_0$ in ATP-depleted HeLa cells indicates solidification due to suppressed metabolic activity. On the other hand, the magnitude of the power-law viscoelasticity $G_1$ remained unchanged compared to untreated cells. Under the severe ATP depletion applied in this study, some cells exhibited blebbing. Severely blebbed cells tend to show decreased $G_1$, likely because the dilution of cytoplasmic contents alleviates crowding. Since it was challenging to completely exclude blebbed cells from MR measurements, the unchanged $G_1$ may reflect an average of hardening due to suppressed metabolism and softening due to blebbing. While details will be reported elsewhere, $G_1$ shows a weak but significant increase under a moderate ATP depletion (0.1 – 1.0 mM sodium azide) that avoids blebbing.

Taken together, these results confirm that, once experimental artifacts are eliminated, cytoplasmic rheology is remarkably invariant – a single power-law across cell types and

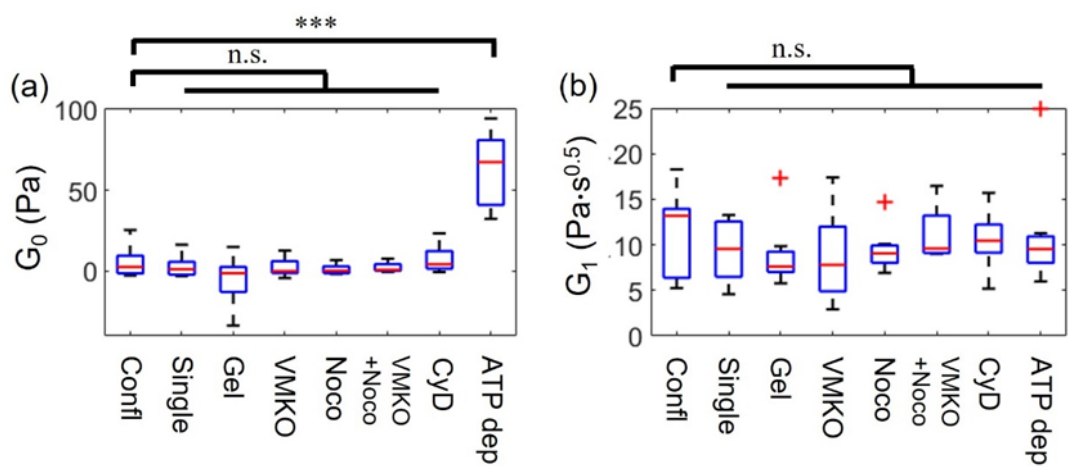

**Figure 7. Quantitative comparison of cytoplasmic viscoelasticity under various conditions in HeLa cells** (a) A magnitude of elasticity plateau $G_0$ at low frequencies and (b) a prefactor $G_1$ for high-frequency viscoelasticities. Complex moduli in Fig. 4 – 6 were fitted by $G = G_0 + G_1(-i\omega)^{0.5}$. The Mann–Whitney U test was used to calculate the P value for differences between confluent HeLa cells and other conditions. ***: P < 0.001. N.S.: P > 0.05. Red plus sign (+) represents an outlier.

conditions – as long as metabolic activity is intact.

**Cytoplasmic rheology of different cell types**

To investigate the dependence of cytoplasmic rheology on cell type, we compared the complex shear modulus of HeLa, Madin-Darby Canine Kidney (MDCK), and embryonic stem (ES) cells (Fig. 8 (a)). As shown in Fig. 8 (b) and 8 (c), the single power-law rheology $G = G_1(-i\omega)^{0.5}$ holds consistently across all these cell types, allowing us to determine $G_1$ by fitting the data. The coefficient $G_1$ varied depending on the cell type, with ES cells exhibiting the lowest viscoelasticity and MDCK cells were slightly stiffer than HeLa cells (Fig. 8 (d)). Additionally, we observed that the viscoelasticity of differentiated ES cells tended to be higher than that of undifferentiated cells (Fig. S3 in supplementary). These findings reveal an intriguing relationship between cell type-dependent rheology and metabolic activity in confluent cell layers. The observed order of viscoelasticity (MDCK > HeLa > ES cells) appears to be inversely correlated with the typical proliferation rates and metabolic activities of these cell types. ES cells, known for their high proliferation rate[50] and metabolic activity[51,52], show the lowest viscoelasticity. This aligns with our earlier observation that ATP depletion leads to increased viscoelasticity, suggesting that higher metabolic activity correlates with lower viscoelasticity.

The intermediate viscoelasticity of confluent HeLa cells may reflect a balance between their high proliferation rate and the need for structural stability in these cancer cells within a monolayer. Confluent MDCK cells, with their epithelial nature and slower proliferation rate, exhibit the highest viscoelasticity, possibly indicating a lower overall metabolic rate compared to HeLa and ES cells. The increase in viscoelasticity observed in differentiated

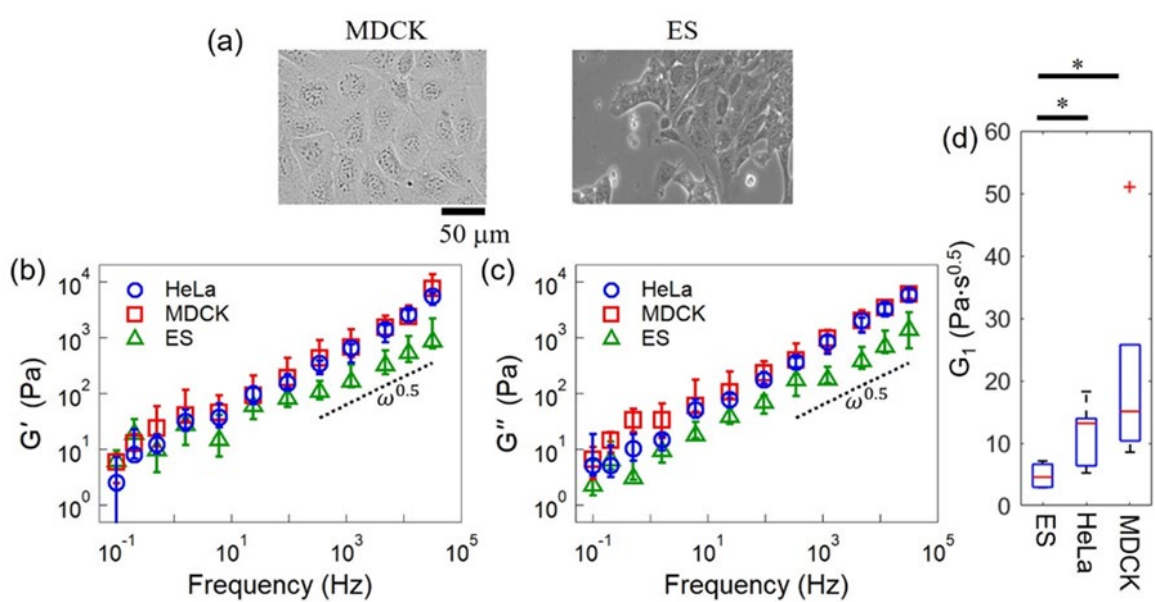

**Figure 8. Cytoplasmic rheology of different cell types**

(a) A microscopic image of MDCK (left) and ES cells (right). (b) $G'(\omega)$ and (c) $G''(\omega)$ of confluent HeLa cells (blue, n = 12), MDCK cells (red, n = 5), and ES cells (green, n = 4). The broken line shows the power law $\propto \omega^{0.5}$. (d) A boxplot of magnitude of viscoelasticity $G_1$. The Mann–Whitney U test was used to calculate the P value for differences between ES cells and other cells. *: P < 0.05. No symbol: P > 0.05. Red plus sign (+) represents an outlier.

ES cells compared to undifferentiated ES cells further supports this relationship. As ES cells differentiate, their metabolic profile typically shifts, often resulting in a reduced overall metabolic activity, which appears to be reflected in their increased viscoelasticity.

**Cytoskeletal structure affects the slow and large scale dynamics**

As shown in previous sections, variations in cytoskeletal development have only marginal effects on the viscoelasticity $G$ of cell cytoplasm measured by AMR. Another important mechanical property of crowded cytoplasm is shear viscosity $\eta$, which can be obtained from the mobility $M$ of a probe particle[43]. In this study, mobility $M$ was measured by applying a constant force to the probes using the feedback technology (see Methods for details). Viscosity $\eta$ of the surrounding cytoplasm was then calculated by using Stokes' equation $M = 1/6\pi\eta a$. Figures 9 (a) and (b) depict the typical trajectories of a probe particle during the particle-pulling measurements in a vimentin-knockout HeLa cell. By applying a very weak force (0.1 – 0.6 pN), the probe particle moved only ~ 1 μm over 1 h of continuous traction. The red and blue trajectories represent the sequential movements of the same probe particle, pulled in opposite directions with positive and negative forces, $F_+$ and $F_-$, respectively.

Even without the application of an optical-trapping force, the probe particles drift due to mesoscopic flows within cells. To eliminate the influence of this drift, we calculated the mobility as $M = (v_+ - v_-)/(F_+ - F_-)$. Here, $v_+$ and $v_-$ were obtained from linear fits to the probe displacements (red and blue curves in Fig. 9 (b)) over time. As shown in Fig. 9 (c), the mobility $M$ of untreated HeLa cells occasionally exhibited negative values, indicating that the probes sometimes moved in the opposite direction to the applied force. This is due to strong drift flows driven by force-generating mechano-enzymes within cells. Cellular migration and shape changes may also contribute to probe movement. These nonequilibrium fluctuations are persistent yet stochastic over a long-term period, with direction potentially reversing on an hourly timescale. Consequently, mobility $M$ exhibited larger variation than the viscoelasticity measured by AMR, reflecting substantial perturbations within the cellular environment. Nonetheless, the statistical average, which is positive, indicates the influence of cytoskeletal development, as described below.

Figures 9 (c) and (d) show the mobility and shear viscosity under the cytoskeletal inhibition. Disrupting the vimentin network resulted in a significant increase in mobility compared to untreated HeLa cells (Fig. 9 (c)). As viscosity is inversely proportional to mobility, vimentin-knockout cells exhibited lower viscosity than untreated cells (Fig. 9 (d)). Next, microtubule networks were inhibited in vimentin-knockout cells by applying nocodazole. Counterintuitively, microtubule disruption led to a moderate decrease in mobility and an increase in viscosity. Thus, shear viscosity is more sensitive to cytoskeletal development than the viscoelasticity measured by AMR.

The effect of substrate stiffness on mobility and shear viscosity was also quantitatively examined (Fig. 9 (c) and (d)). In HeLa cells on soft gels, $M$ and $\eta$ were intermediate between those of untreated and vimentin-knockout cells, and they were similar to values observed in cells with simultaneous disruption of the vimentin and microtubule networks. On soft gels, the development of actin and microtubule networks is strongly inhibited,

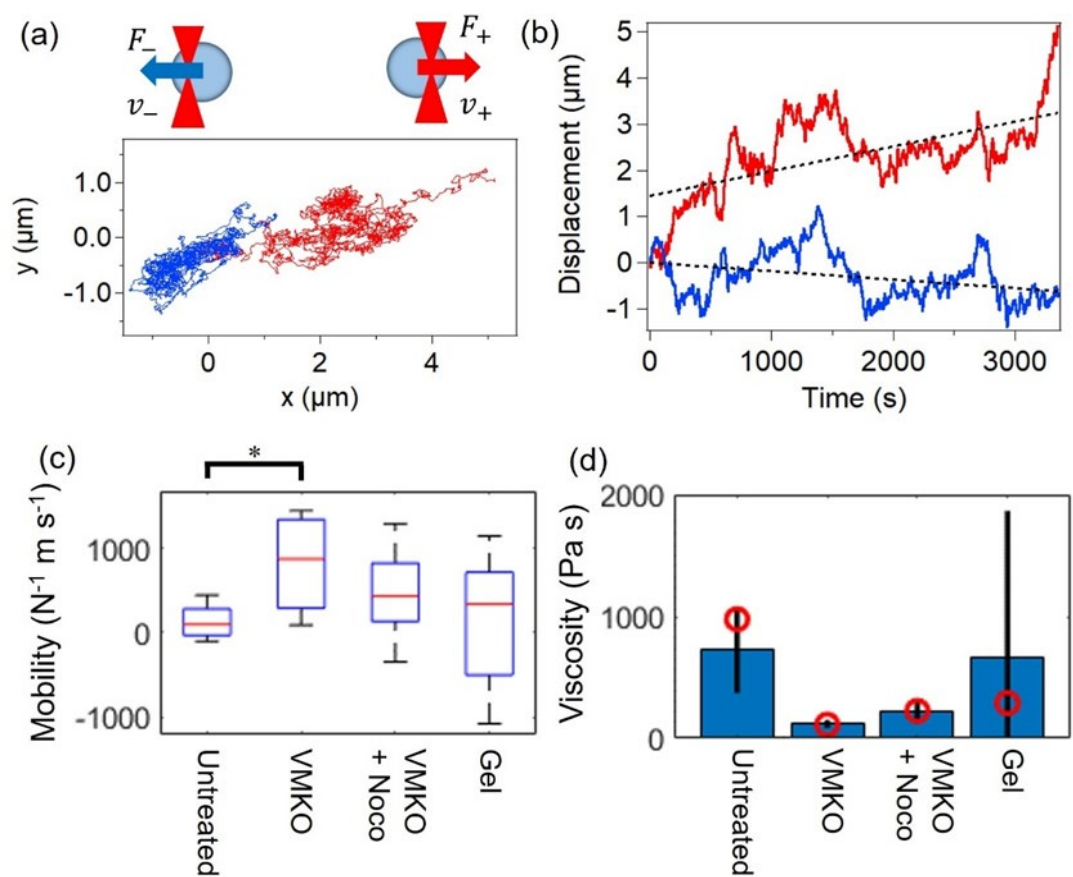

**Figure 9. Cytoskeletal structure affects the slow and large scale dynamics**

(a) Typical probe trajectories under constant force application in the $\pm x$ directions in a vimentin-knockout HeLa cell. A force was initially applied in the negative $x$-direction (blue), pulling the probe particle for 1 h. Subsequently, the same probe particle was pulled in the positive $x$-direction (red). For clarity, the origin of both trajectories (blue and red) was set to the point where traction was initiated. (b) Time series of particle displacement in the direction of the applied force. In both (a) and (b), the blue trajectory/curve represents the condition under negative force ($F_-$), and the red trajectory/curve represents the condition under positive force ($F_+$). Dependence of (c) the mobility, $M$, and (d) shear viscosity $\eta$ on cytoskeletal inhibition. (d) The bar graph shows the viscosity calculated from the average mobility, $\eta = 1/6\pi a \langle M \rangle$. Error bars indicate standard errors, calculated by error propagation. Red circles represent the viscosity calculated using the median value of mobility. (c, d) Untreated HeLa cells (n = 9). VMKO HeLa cells (n = 6). VMKO+Noco HeLa cells (n = 11). HeLa cells on soft gel (n = 8). *: P < 0.05.

while a sparse vimentin network remains[47]. Thus, alterations in the cytoskeletal network due to the substrate stiffness may affect the shear viscosity of cells.

Figures 9 (c) and (d) indicate that, unlike the viscoelasticity measured by AMR, shear viscosity is significantly influenced by the vimentin network, but not by the microtubule and actin networks. This distinction underscores the specific role of the vimentin network in regulating the large-scale mechanical properties of the cytoplasm [17].

**DISCUSSION**

The viscoelasticity of the cell interior measured in this study consistently exhibited a single-exponent power-law rheology of $G = G_1(-i\omega)^{0.5}$, regardless of the cell type, micromechanical environment, and cytoskeletal structure. These characteristics of cytoplasmic viscoelasticity (Fig. 3-6) markedly differ from previous studies, which reported a viscoelasticity characterized by $G = C_1(-i\omega)^a + C_2(-i\omega)^b$ with $0 \lesssim a < 0.3$ and $b \sim 0.75$. This viscoelastic behavior, especially $b \sim 0.75$, is a hallmark for the semiflexible biopolymer networks, a well-known theoretical model for cytoskeletons[53,54]. In many of these prior studies, probe particles were introduced into cells via phagocytosis[11,55], resulting in strong interactions with the cytoskeleton (Fig. 10 (a)). The lipid plasma membranes that covered the probe particles during phagocytosis acted as a scaffold for the cytoskeleton. Consequently, cytoskeletal inhibition significantly reduced the measured viscoelasticity[11]. We also confirmed that the viscoelasticity differs fundamentally from $G = G_1(-i\omega)^{0.5}$ when a probe particle strongly interacts with an actin cortex (see supplementary, Fig. S4). In these cases, viscoelasticity shows broad dispersions depending on the probe's location, whether at the cell membrane or within the cell[14], reflecting the heterogeneity of the cytoskeletal structure and the probe-cytoskeleton coupling. In this study, the surfaces of the particles were passivated with polyethylene glycol (PEG) strands, preventing adhesion to the cytoskeleton[56], explaining why cytoskeletal development had only a minor effect on the viscoelasticity measured in this study (Fig. 7 (b)).

Although PEG-modified probe particles were used in some previous AMR studies in cells, those studies employed much stronger laser power (tens of mW), inducing much larger displacements to probes (~100 nm)[36] than in this study (~ 0.5 mW). When 1 μm probe particle undergoes 100 nm displacement, more than 10% of the strain is generated around the probe surface, which is sufficient to induce nonlinear responses of biological materials such as cytoskeleton[57]. In contrast, our AMR measurements used much smaller displacement amplitudes, ranging from 0.02 nm to 5 nm depending on the frequency[37], with less than 1 mW of laser power. This is a key feature of our study, enabling us to measure the linear viscoelasticity of the cytoplasm in a minimally invasive manner. We demonstrated that a strong laser power consistently causes cytoplasmic hardening, as shown in Fig. 3 (b). Therefore, the pronounced elasticity plateau observed in previous studies may be attributable to a phototoxic response.

Previous studies using AFM and magnetic twisting cytometry reported significant cell softening upon cytoskeletal inhibition, because these techniques predominantly measure the rheology of the actin cortex at the cell surface (Fig. 10 (a))[20,58,59]. In contrast, we showed that disrupting the actin cytoskeleton does not significantly alter the viscoelasticity of the cell cytoplasm[37]. In this study, we demonstrated that inhibition of other cytoskeletal networks, such as vimentin and microtubules, also had limited effects on the measured viscoelasticity. As discussed, the minor impact of cytoskeletal inhibition on the viscoelasticity of crowded cytoplasm may stem from the weak interaction between the probe and the cytoskeleton.

Suppression of actin and microtubule networks hardly affected particle-pulling MR. In the particle-pulling measurements, the timescale of pulling, $a/V$, which is the inverse of the effective strain rate induced by probe movement, was ranged from $10^3$ to $10^4$ s. Over such long timescales, actin and microtubules can depolymerize and reorganize their structures[60,61]. Furthermore, actin and microtubule cytoskeletons are sparsely expressed in the deeper regions of the cell[47,48,62,63], explaining our particle-pulling MR results. On the other hand, disruption of vimentin significantly reduced shear viscosity. The vimentin network, which is much more stable than actin and microtubule networks within cells[64], is expressed inside the cells, particularly around the nucleus[47,65]. Therefore, it significantly constrains the diffusion of particles and vesicles[17], and hinders the slow unidirectional

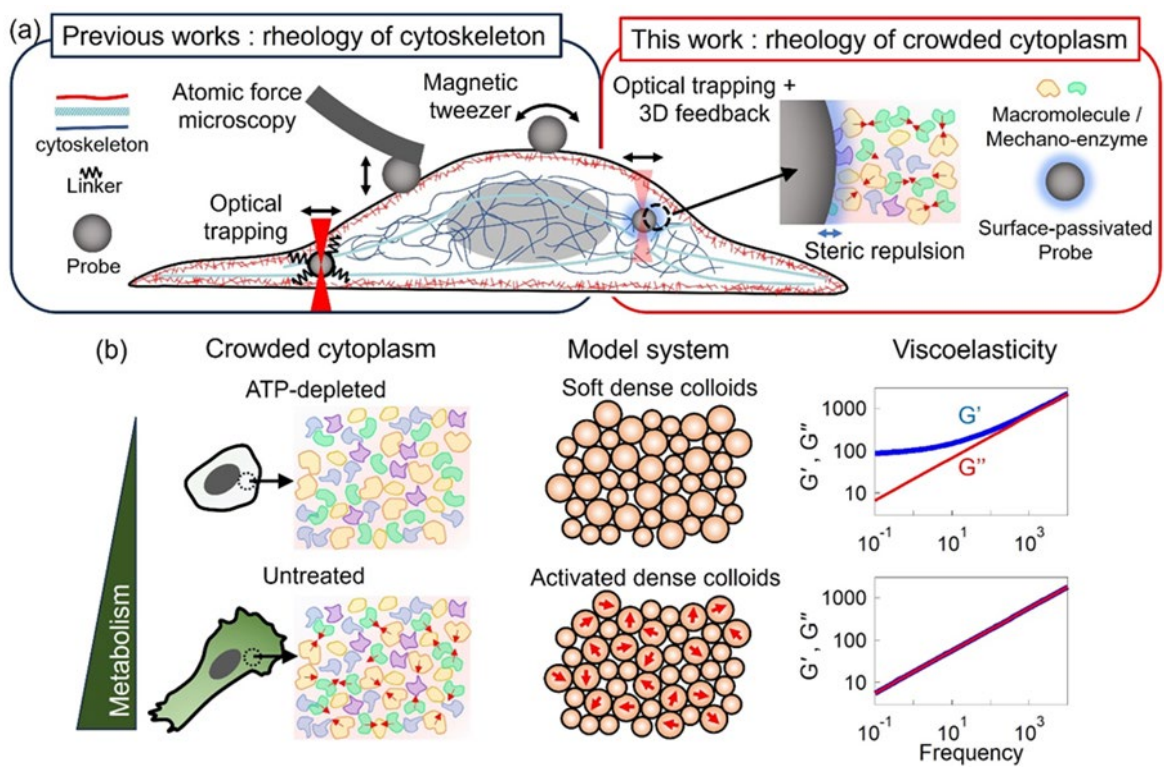

**Figure 10. Schematic illustration of cell rheology**

(a) Previous optical trapping-based MR studies employed strong laser power and phagocytosed probe particles, which strongly interacted with the cytoskeleton of cells. These studies primarily measured the rheology of cytoskeleton inside cells. Other methods, such as magnetic twisting cytometry and atomic force microscopy, predominantly probed the viscoelasticity of the actin cortex in cells, providing results heavily influenced by cytoskeletal development. In this work, optical-trapping-based MR was performed with a weak laser power, with the advantage of a 3D feedback system, which allowed to avoid photodamage. Furthermore, the probe surface was passivated to prevent attachment to cytoskeletal structures. This approach enabled us to measure the rheology of the cytoplasm crowded with macromolecules, revealing properties that are largely independent of cytoskeletal development. (b) The viscoelasticity of soft dense colloids, such as dense nano-emulsions, universally follows the form $G = G_0 + G_1(-i\omega)^{0.5}$, making them a potential model system for ATP-depleted cytoplasm (Fig. 6). In a similar manner to the cytoplasm in untreated cells, mechanically activated dense colloids exhibit a power-law rheology of $G = G_1(-i\omega)^{0.5}$, which may represent a characteristic rheological behavior of dense colloids driven far from equilibrium.

motion of the probe particle as shown in Fig. 9 (c).

Cells are known to adapt their mechanical properties to match those of their surrounding environment by regulating cytoskeletal structures[66]. On softer substrates, cells adapt to a more fluid-like state with a sparse cytoskeleton, whereas on harder substrates, they become more solid-like with denser cytoskeletal structures[67,68]. Similarly, the cell population density influences cytoskeletal development[49], which may in turn affect cellular elasticity[69]. It should be noted that these past findings were made using methods that primarily probe mechanical properties of cytoskeleton. In contrast, our study revealed that the rheology of the crowded cytoplasm did not significantly depend on the micro-mechanical environment, as long as intracellular metabolism was maintained. Since the primary cellular response to the mechanical environment involves adjusting cytoskeletal architecture, our findings, showing a limited impact of both cytoskeletal disruption and mechanical environment on cytoplasmic viscoelasticity, are consistent with the concept of mechanical adaptation in cells.

Our findings imply that the cytoplasm behaves as a typical complex medium that becomes less viscoelastic when stirred. Such nonequilibrium phenomena have been studied in model systems, such as dense colloidal suspensions[70]. The solidification observed in ATP-depleted cells (Fig. 6 (a)) may represent the reverse effect of fluidization, indicating that the nonequilibrium fluctuations induced by metabolic activity fluidize the cytoplasm in normal cells[34,71]. Although the direct impact of cytoskeletal structures on cytoplasmic viscoelasticity is limited, they can indirectly influence it by stirring the cytoplasm. Motor proteins such as kinesin, dynein, and myosin interact with the cytoskeleton and generate vigorous nonequilibrium fluctuations inside the cell[11,36]. When the activities of motor proteins are impaired due to cytoskeletal disruption, these nonequilibrium fluctuations decrease[11,36], potentially increasing the viscoelasticity of the cytoplasm. In our experiments, disrupting microtubules in vimentin-depleted cells caused a small increase in viscoelasticity and shear viscosity (Fig. 7 (b) and 9 (d)). We speculate this is because microtubule motors (kinesin, dynein) normally contribute to metabolic stirring of the cytoplasm; removing them reduces non-equilibrium fluctuations, leading to slight re-solidification. Thus, while the cytoskeleton does not hinder the small oscillations of PEG-coated colloidal particles, it could facilitate cytoplasmic fluidization by inducing intracellular flows on meso- to macroscopic scales.

The power-law behavior observed in cytoplasmic viscoelasticity (Fig. (3) – (8)) aligns with a previous study that measured the fluctuations of 5 nm gold nanoparticles in cells using fluorescence correlation spectroscopy[25]. Despite the considerable difference in probe size, power-law exponents between $10^2$ and $10^5$ Hz were found to be $\sim 0.5$ across 10 different cell types. Assuming that probe fluctuations in such frequencies are predominantly thermal, the results are consistent with our AMR data (Fig. 8). This consistency suggests that the single-power law rheology is universal among mammalian cells, while the coefficient $G_1$ showed about a 2- to 3-fold variation depending on conditions (Fig. 8 (d) and Ref. [25]). The cause of the variability in $G_1$ remains unclear but may be influenced by cell type, cytoplasmic contents and magnitude of metabolic activity. On the other hand, the exponent 0.5 likely originates from a common fundamental property of the cell cytoplasm, as the thermal fluctuations of the nanoparticles would primarily reflect the mechanics of the interstitial medium between the cytoskeletal networks. Ineffectiveness of cytoskeletal inhibition on cytoplasmic viscoelasticity in our AMR further supports this interpretation.

Regardless of cell type, the cell cytoplasm harbors a dense population of macromolecules, constituting 20 – 40 % of the total volume of the cell[72-74]. Additionally, it suspends mesoscopic colloidal components, including organelles and condensates formed via phase separation[75,76]. Thus, the mechanical properties of the cytoplasm, filling the interstices of cytoskeletal network, behave akin to a concentrated emulsion or colloidal suspension[41,77]. For example, viscosity of cytoplasm exponentially increases with the concentration of intracellular solid contents due to osmotic compression, resembling the behavior of glass formers[41,77]. Additionally, viscosity of metabolically inert cell extracts almost diverges around physiological concentrations[77], akin to dense colloidal suspensions near the glass transition point[78].

Similarity between cell cytoplasm and colloidal glass suggests that the single exponent 0.5 observed in cytoplasmic viscoelasticity could be related to the rheology of densely packed colloids[42]. When soft colloids reach a high packing density, they undergo solidification above a critical volume fraction known as the jamming transition. Above the jamming packing fraction, soft colloids demonstrate anomalous viscous loss, with $G' \propto \omega^0$ and $G'' \propto$ $\omega^{0.5}$, at low frequencies[79]. This anomalous viscous loss is commonly observed in amorphous solids, reflecting the marginal stability in soft, jammed solids[80] (Fig. 10 (b)). Given the intracellular molecular crowding[72], this viscous loss may contribute to the observed rheology as $G = G_0 + G_1(-i\omega)^{0.5}$ in ATP-depleted HeLa cells. In contrast, the elasticity plateau $G_0$ disappears in metabolically active cells, suggesting that non-thermal fluctuations due to metabolic activity alter cytoplasmic rheology. The power law rheology $G \propto (-i\omega)^{0.5}$ is also observed in mechanically activated dense systems, such as vibrated dense granular media[81], sheared concentrated starlike micelles[82], and dense motile bacterial suspensions[83] (Fig. 10 (b)). While the specific mechanism behind this rheological behavior requires further numerical and theoretical investigation, the power-law rheology with an exponent of 0.5 could be a feature of dense colloids far from equilibrium.

This study examined the dynamic mechanical properties of crowded cytoplasm under diverse conditions, including cytoskeletal inhibition, different mechanical environments, and across multiple cell types. Using optical trapping with a 3D feedback system, we performed microrheology with a lower laser power. By eliminating photodamage and probe-cytoskeleton coupling, we revealed the inherent mechanical behavior of cytoplasm. Remarkably, the complex modulus of cell cytoplasm consistently exhibited a single-exponent power-law rheology of $G = G_1(-i\omega)^{0.5}$, over a vast frequency range ($10^{-1}$ to $10^5$ Hz) as long as metabolic activity was maintained. Only ATP depletion disrupted this behavior, introducing a low-frequency plateau $G_0$ and significantly solidifying the cytoplasm. This universal behavior enables quantitative comparison of intracellular mechanics across different cell states (e.g., healthy vs diseased cells, differentiated vs stem cells, etc.) using just two parameters ($G_1$ and $G_0$). While cytoskeletal components like vimentin networks suppressed the long-term and large-scale transport in response to an infinitesimal external force, their limited impact on the power-law behavior suggests that the universal rheology arises from macromolecular crowding, a conserved feature of mammalian cells[25]. Overall, our results indicate that mammalian cells maintain a soft glassy cytoplasmic matrix through metabolic activity. While the unified physical description of living cytoplasm qualitatively explains both the universality of the ~0.5 exponent and the changes under metabolic inhibition, the precise molecular mechanism underlying the universality remains to be explored in the future.

## MATERIALS AND METHODS

### Cell culture

HeLa, vimentin-knockout HeLa, and Madin-Darby Canine Kidney (MDCK) cells were cultured in Dulbecco's modified Eagle's medium (D-MEM, high glucose; Wako) supplemented with glucose (1 mg/ml), penicillin (100 U/ml), streptomycin (0.1 mg/ml), and 10% fetal bovine serum (FBS) at 37 °C in a humidified atmosphere containing 5% $CO_2$. For vimentin-knockout HeLa cells, 40 μg/ml puromycin (Invitrogen) was added to this medium[84].

HeLa and vimentin-knockout HeLa cells were seeded on fibronectin-coated glass-bottom dishes or fibronectin-modified polyacrylamide (PAA) gels whith a thickness of ~ 50 μm. MDCK cells were seeded on glass-bottom dishes coated with collagen gel (CellMatrix Type I-A; Nitta Gelatin Inc.) with a thickness of ~ 50 μm. All cells were incubated overnight before the introduction of probe particles.

Undifferentiated EB5 cells derived from mouse embryonic stem (ES) cells were cultured on i-Matrix in Glasgow Minimum Essential Medium (G-MEM; Wako) supplemented with 10% fetal clone serum (FCS; GE Healthcare Life Sciences), 1x nonessential amino acids (NEAA) solution (Nacalai Tesque), 1 mM sodium pyruvate (Nacalai Tesque), 100 mM 2-mercaptoethanol (2-ME; Nacalai Tesque), and 1000 U/ml leukemia inhibitory factor (LIF; Wako) at 37 °C in a humidified atmosphere containing 5% $CO_2$. Before introducing probe particles for MR experiments, these cells were seeded on i-Matrix-coated glass-bottom dishes and incubated overnight.

## Cell preparation for microrheology

The surface of the spherical probe particles (melamine particles, 1 μm diameter; microParticles GmbH) was coated with polyethylene glycol (PEG) strands (mPEG-NH2, 1000Da, PG1-AM-1k; NANOCS)[85] to passivate the probe particle surface. In aqueous environments, hydrophilic PEG acts as a polymer brush, preventing adhesion to other objects or molecules. Probe particles were introduced into cells using a gene gun (PDS-1000/He; Bio-Rad). Excess beads that did not enter the cells were removed by washing the dishes with phosphate-buffered saline (PBS). After replenishing the dishes with fresh culture medium, they were placed in a $CO_2$ incubator overnight to allow the cells to recover from any potential damage.

Since the introduction of $CO_2$ into the experimental chamber adds noise to the signal, MR experiments were conducted in $CO_2$-independent culture medium (L-15 medium; Gibco) supplemented with 10% FBS. All measurements were performed at 34 °C. For MR experiments under the condition of ATP depletion, FBS was omitted from the medium as specified below.

## Fabrication of polyacrylamide (PAA) gels

PAA gels were prepared based on previously published protocols with minor modifications[86] (see Supplementary Materials for details). Briefly, glass-bottom dishes were amino-silanated using 3-Aminopropyltriethoxysilane and glutaraldehyde. A mixture of acrylamide, bis-acrylamide, N,N,N',N'-Tetramethylethylenediamine, and ammonium persulfate was sandwiched with the amino-silanated dish and a plasma-cleaned cover glass. After gelation, the PAA gel surface was functionalized by sulfo-SANPAH under UV light[87] and subsequently coated with fibronectin. The resulting PAA gel was approximately 50 μm thick. The elasticity was measured with passive MR.

## Cytoskeletal disruption

To study cells lacking a vimentin network, vimentin-knockout HeLa cells were generated by transfection with a human vimentin CRISPR/Cas9-knockout plasmid and a human vimentin homology-directed DNA repair plasmid[84] (Santa Cruz Biotechnology). The actin and microtubule networks in HeLa cells were disrupted by adding 2.5 μg/ml cytochalasin D (Sigma) and 3 μg/ml nocodazole (Sigma), respectively. Before performing MR experiments, the culture medium was replaced with a cytoskeleton disruption medium (cytochalasin D or nocodazole in L-15 medium with 10% FBS). The cells were then incubated at 34 °C on the MR setup for approximately 60 minutes to ensure that the actin and microtubule networks were fully disrupted.

## ATP depletion

HeLa cells were incubated in ATP depletion medium (L-15 medium with 50 mM 2-deoxy-D-glucose(Carbosynth) and 10 mM sodium azide (Sigma), without FBS[7]) at 37 °C for approximately 12 hours prior to the MR measurements to ensure that ATP stored in the cells was fully consumed[37].

**AMR in living cells**

We briefly summarize the key principles of feedback active microrheology (AMR)[14,37]; see the supplementary materials for details. The displacement of a probe particle inside the cell was measured using a fixed probe laser. The laser focus was dynamically adjusted to track the center of the strongly fluctuating probe particle via feedback control of the sample stage position. Simultaneously, a small sinusoidal force was applied to the probe particle using a drive laser. From the applied force and resulting displacement, the complex shear modulus $G(\omega)$ was calculated using the generalized Stokes relation. Importantly, because the feedback system eliminates the need to optically trap the particle, AMR can be performed using minimal laser power, thereby reducing phototoxic effects.

**Particle-pulling experiment**

To measure the shear viscosity $\eta$, a constant force was applied to a probe particle using a single-drive laser combined with the stage feedback technology. The total displacement of the probe particle, $u(t)$, was determined by summing the laser position and the displacement of the sample stage. A linear fit of $u(t)$ over time $t$ yielded the particle velocity $v$, from which the mobility $M$ and shear viscosity $\eta$ were calculated as:

$$M = v/F, \quad \eta = F/6\pi a v.$$

**Statistical analysis**

Since the distributions of $G'$ and $G''$ were non-Gaussian, statistical analyses were performed using the Mann–Whitney U test, unless otherwise stated in the Figure captions. $P$ values < 0.05 were considered indicative of a significant difference. All statistical tests were conducted using MATLAB software.

**ACKNOWLEDGEMENTS**

This work was supported by JSPS KAKENHI Grant Number JP 24K00601, JP24K21535, JP22K03552, JP25K00220, and the JSPS Core-to-Core Program "Advanced core-to-core network for the physics of self-organizing active matter" (JPJSCCA20230002). We thank Hitoshi Niwa in Kumamoto University, Yujiro Sugino, and Wataru Nagao in Kyushu University for helpful discussions and technical support.

**CONFLICTS OF INTEREST** The authors declare no conflict of interest.

**AUTHOR CONTRIBUTION**

H. E., K. N., F.A.S. v. E., Y. T., and S. I. collected/analyzed MR data. H. I. provided the samples. H. E., and D. M. analyzed/discussed the results. D. M. designed/supervised the project. H. E. and D. M. wrote the manuscript. All authors have read the manuscript, agreed to its content, and approved the submission.